\documentclass[prd,aps,superscriptaddress,floatfix,nofootinbib,showpacs,twocolumn]{revtex4}
\usepackage{exscale}                  
\usepackage[intlimits]{amsmath}       
\usepackage{amsfonts}
\usepackage{amssymb,amscd}
\usepackage[dvips]{epsfig}                   
\usepackage{array}
\usepackage{wrapfig}
\usepackage{color}
\newcommand{\beqa}{\begin{eqnarray}}
\newcommand{\eeqa}{\end{eqnarray}}
\newcommand{\beq}{\begin{equation}}
\newcommand{\eeq}{\end{equation}}
\newcommand{\gS}[1]{#1\!\!\!\!\!\not~}	
\newcommand{\GS}[1]{#1\!\!\!\!\!\!\!\not~}

\newcommand{\pslash}{\gS{p}}
\newcommand{\Pslash}{\GS{P}}

\newcommand{\al}{\alpha}
\newcommand{\ba}{\beta}
\newcommand{\ga}{\gamma}
\newcommand{\de}{\delta}

\newcommand{\G}{\Gamma}

\newcommand{\tr}{\textrm{tr}}

\def\Eq#1{Eq.~(\ref{#1})}

\newcommand{\lqcdsq}{\Lambda^2_{\mathrm{QCD}}}

\begin{document}

\title{Volume behaviour of quark condensate, pion mass and decay constant 
from Dyson-Schwinger equations}
\author{Jan Luecker}
\affiliation{Institut f\"ur Kernphysik, 
  Technische Universit\"at Darmstadt,
  Schlossgartenstra{\ss}e 9,\\ 
  D-64289 Darmstadt, Germany}
\author{Christian~S.~Fischer}
\affiliation{Institut f\"ur Kernphysik, 
  Technische Universit\"at Darmstadt,
  Schlossgartenstra{\ss}e 9,\\ 
  D-64289 Darmstadt, Germany}
\affiliation{GSI Helmholtzzentrum f\"ur Schwerionenforschung GmbH, 
  Planckstr. 1  D-64291 Darmstadt, Germany.}
\author{Richard Williams}
\affiliation{Institut f\"ur Kernphysik, 
  Technische Universit\"at Darmstadt,
  Schlossgartenstra{\ss}e 9,\\ 
  D-64289 Darmstadt, Germany}
\date{\today}

\begin{abstract}
We solve the coupled system of Dyson-Schwinger and Bethe-Salpeter 
equations for the quark propagator and the pion Bethe-Salpeter 
amplitude on a finite volume. To this end we use a truncation 
scheme that includes pion cloud effects in the quark propagator
and light mesons. We study volume effects in the quark condensate, 
the pion mass and the pion decay constant and compare to corresponding
results in other approaches. In general we find large effects
for volumes below $V=(1.8 \, \mbox{fm})^4$.
\end{abstract}


\maketitle


\section{Introduction}\label{sec:intro}

Dynamical chiral symmetry breaking is one of the fundamental properties 
of QCD. Its pattern determines the experimentally observable spectrum of 
light hadrons as well as the details of the underlying interaction between
quarks and gluons. It therefore plays a central role in our understanding 
of QCD. Despite great efforts in the last decades not all details of 
dynamical chiral symmetry breaking are as well understood as one could wish. 

In this respect, the chiral properties of QCD at finite volume have attracted
considerable attention over the years. Strictly speaking, dynamical chiral 
symmetry breaking and the associated formation of Goldstone bosons is 
restricted to the infinite volume limit. On any given finite volume the 
corresponding order parameter, the quark condensate $\langle \bar{q}q\rangle$, 
goes to zero in the limit of vanishing bare quark mass $m$. 
Nevertheless, one can extract 
the properties of the infinite volume theory from the formulation in a box.
For large enough volumes and/or quark masses the box effects are small
and the finite volume theory closely resembles its infinite volume limit.
This happens if the condition  
\beq
m \, V \, \langle \overline{q}q\rangle \gg 1 \,,
\eeq
is satisfied and the eigenvalues of the Dirac operator are almost dense
\cite{Leutwyler:1992yt}. On the other hand, if 
$m \, V \, \langle \overline{q}q\rangle \ll 1$ 
dynamical chiral symmetry breaking is lost. 

Another important condition is the relation
of the four-volume $V$ with a typical hadronic scale $\Lambda_H$ and the 
pseudo-Goldstone mass $M_\pi$. If the inequality 
\beq
\frac{1}{\Lambda_{\mathrm{H}}^4} \ll V \ll \frac{1}{M_\pi^4} \,,
\eeq
is satisfied, the theory is in a critical region where long range correlations
are at work. Here the QCD partition function depends on the quark mass $m$
and the volume only through the combination $\mu := m \, V \, \Sigma$, where
$\Sigma$ is the chiral condensate in the infinite volume limit. In this critical
region exact, 
analytic and universal scaling laws are known from chiral random matrix 
theory \cite{Shuryak:1992pi,Verbaarschot:1993pm,Akemann:1996vr}, see {\it e.g.}
\cite{Verbaarschot:2009jz} for a recent review. Certainly these are of great 
importance for approaches like lattice QCD which are inherently limited to 
finite volumes and lattice spacings.
\begin{figure*}[t]
\centerline{\includegraphics[width=1.4\columnwidth]{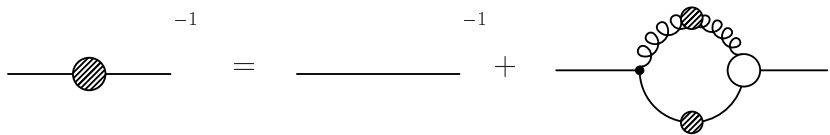}}
\caption{The Dyson-Schwinger equation for the quark propagator. Filled circles
denote dressed propagators whereas the empty circle stands for the dressed
quark-gluon vertex.}
\label{DSEs}
\end{figure*}

The properties of QCD at finite volumes have been studied by several approaches.
One, of course, is lattice QCD, see 
\cite{Aoki:1993gi,Damgaard:1999tk,Guagnelli:2004ww,Orth:2005kq,Fukaya:2007fb,Necco:2009cq}
and Refs. therein, but important insights have also been obtained
from effective theories such as chiral perturbation theory 
\cite{Gasser:1986vb,Colangelo:2005gd,Damgaard:2008zs},  
or the quark-meson model \cite{Berges:1997eu,Braun:2004yk,Braun:2005gy}.
In this work we follow a complementary approach. We study volume effects in the 
quark condensate, the pion mass and the pion decay constant from the framework 
of Dyson-Schwinger equations (DSEs) of the quark propagator and the corresponding 
Bethe-Salpeter equations (BSEs) for pseudoscalar bound state 
amplitudes. 

One of the advantages of this approach is that volume effects can be studied 
continuously from very small to very large volumes (corresponding studies for 
meson observables using chiral perturbation theory for example have to distinguish 
between two different regions of chiral counting). The implementation of mixed 
boundary conditions in the spatial and time directions is possible without great 
effort. Furthermore one has direct access to the infinite volume and the continuum 
limit without the need to perform any extrapolations. We are thus in a position 
to study chiral symmetry restoration at small volumes together with effects at 
large and infinite volumes in the same framework. 

On the other hand we need to work with a truncation scheme in order to close 
our system of equations. This scheme is systematic in the following sense: 
in a first step we neglect the volume dependence of the three-point functions.
For the quark-gluon vertex we work with a volume independent ansatz that has 
proven useful in calculations of pion cloud effects in the masses of light 
mesons \cite{Alkofer:2008et,Fischer:2008wy}. The volume dependence of the gluon
and quark propagators are then determined from their Dyson-Schwinger equations
and serve as input into the volume dependent Bethe-Salpeter equation of the
pion. This truncation scheme can be systematically extended to also include 
volume effects of higher Green's functions by explicitly solving 
their corresponding DSEs.

This work is organised as follows: In the next two subsections we summarise the 
properties of the quark Dyson-Schwinger equation and the Bethe-Salpeter equations 
on a torus. We then discuss details of the numerical treatment of these equations 
in subsection \ref{sec:num}, whereas in subsection \ref{sec:trunc} we specify our 
approximation scheme for the quark-gluon vertex. Our numerical results for the 
quark condensate are presented in section \ref{num:quark}. We then proceed to 
discuss the volume dependence of the pion mass and decay constant in subsection 
\ref{num:pi}. Starting with subsection \ref{num:gluon} we also take into account 
finite volume effects in the gluon propagator.
We conclude and summarise in section \ref{sum}.

\section{Quarks and pions at finite volume}\label{sec:finvol}
\subsection{The quark Dyson-Schwinger equation on a torus}\label{sec:dse}

In Euclidean momentum space, the renormalised dressed gluon and 
quark propagators in the Landau gauge are given by
\beqa
D_{\mu \nu}(p)    &=&   \left(\delta_{\mu \nu} -\frac{p_\mu p_\nu}{p^2}\right) 
                        \frac{Z(p^2)}{p^2} \,, \label{gluon}\\
S(p)              &=& \frac{1}{i \pslash A(p^2) + B(p^2)}\,.  \label{quark}
\eeqa
Here the gluon dressing function $Z(p^2)$ and the quark wave function 
$Z_f(p^2)=A^{-1}(p^2)$ also depend on the renormalisation point, whereas the quark 
mass function $M(p^2)=B(p^2)/A(p^2)$ is a renormalisation group invariant. These propagators 
can be calculated from their Dyson-Schwinger equations (DSEs). Most 
important for this work is the DSE for the quark propagator which is shown 
diagrammatically in Fig.~\ref{DSEs}. 

On a compact manifold, the gluon and quark fields have to obey appropriate
boundary conditions in the time direction. These have to be periodic for
the gluon fields and antiperiodic for the quarks. It is convenient, though 
not necessary, to choose the same conditions in the spatial directions. 
For the gluon propagator we will use periodic boundary conditions in all four
space-time directions. For the quark field we will employ antiperiodic 
boundary conditions in all four directions. In principle it is also possible
to implement periodic boundary conditions for the spatial directions of the
quarks; corresponding results will be presented elsewhere.

We choose the box to be of equal length $L$ in all spatial 
directions and a potentially longer value $T$ in the time direction. The 
corresponding volume is denoted $V=T\,L^3$.  
Together with the boundary conditions the finite volume leads to 
discretised values in momentum space. Thus all momentum integrals 
appearing in the Dyson-Schwinger equations are replaced by sums over Matsubara 
modes. For the ghost and gluon DSE the corresponding equations as well as their
solutions have been discussed in detail in 
Refs.~\cite{Fischer:2002eq,Fischer:2005ui,Fischer:2007pf}.
The corresponding solutions in the infinite volume/continuum limit are given
and discussed in Ref.~\cite{Fischer:2008uz}. To make this paper self-contained 
we shortly repeat the general features of these solutions below, where we 
discuss the quark-gluon interaction. 

The quark-DSE on a torus has been discussed in Ref.~\cite{Fischer:2005nf,Fischer:2007ea} 
at zero and in Ref.~\cite{Fischer:2009wc} at finite temperature, where results for 
the quark propagator and condensate at real quark momenta have been given. 
In the present work the quark propagator also serves as input into the 
Bethe-Salpeter equation for pseudoscalar mesons. As we will see this 
also necessitates the determination of the quark propagator for complex momenta. 

Restricting our system to a finite volume results in the Matsubara modes
\begin{equation}
p_{i} = \frac{2\pi}{L_i}n_i; \,\,\,\,\,\,\, p_{i} = \frac{2\pi}{L_i}(n_i+1/2)
\end{equation}
for periodic and antiperiodic boundary conditions in direction $i$ with 
length $L_i$ and $n_i \in \mathbb{Z}$. With discretised momenta an integral 
in momentum space becomes
\begin{equation}
\int \frac{d^4q}{(2\pi)^4} \rightarrow \frac{1}{L^3 T} \sum_{n_1,n_2,n_3,n_4=-N}^{N_{max}}\,,
\end{equation}
where $N$ defines the maximal Matsubara mode, and therefore acts as an ultraviolet cutoff.
Furthermore $N_{max}=N$ for periodic and $N_{max}=N-1$ for antiperiodic boundary conditions.
Similar to the infinite volume case where one uses hyperspherical coordinates we
rearrange the summation over Matsubara frequencies in the following convenient way
\begin{equation}
\sum_{n_1,n_2,n_3,n_4=-N}^{N_{max}} = \sum_{i}\sum_{m}\,,
\end{equation}
where $i$ counts hyperspheres containing all momenta with the same absolute value
and $m$ counts individual momenta on each sphere. In the following we frequently
use the notation 
\beq
\sum_j := \sum_{i}\sum_{m}\,,
\eeq
to account for the sum over all 
momenta $q_j$.
\begin{figure}[t]
\centering
\includegraphics[width=0.4\linewidth]{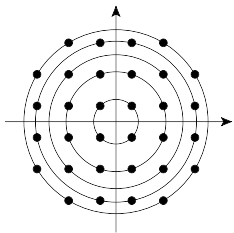} 
\caption{The discretised momentum space that will be used:
 the momenta for antiperiodic boundary conditions in all directions. 
 The circles indicate that only `full' hyperspheres are taken into 
 account, see text for details. This procedure leads to reduced
 cutoff effects. \label{fig:Tori}}
\end{figure}

Fig.~\ref{fig:Tori} shows how the discretised momentum space looks in two dimensions,
with the circles indicating the hyperspheres that are counted by $i$. The hyperspheres 
that are not fully occupied for a given value of $N$ are not used, {\it i.e.} 
hyperspheres that receive additional vectors when going from $N$ to $N+1$ are discarded. 
This procedure leads to an approximate restoration of rotational symmetry for large momenta.
Consequently cutoff artefacts in the ultraviolet momentum regions are drastically 
reduced \cite{Fischer:2002eq,Fischer:2005nf,Goecke:2008zh}. 

The Dyson-Schwinger equation for the quark propagator on a torus is then given by
\begin{widetext}
\beq
S^{-1}(p_{i}) \;=\; 
Z_2 \, [S^0(p_{i})]^{-1} -  
C_F\, \frac{Z_2}{\widetilde{Z}_3}\, \frac{g^2}{L^3T} \sum_{j} \,
\gamma_{\mu}\, S(q_{j}) \,\Gamma_\nu(q_{j},p_{i}) 
\,D_{\mu \nu}(p_{i}-q_{j}) \,. \label{quarkDSE}
\eeq
\end{widetext}
where the factor $C_F = 4/3$ stems from the colour trace and we have introduced 
a reduced quark-gluon vertex $\Gamma_\nu(q,p)$, by defining
$\Gamma^{full}_{\nu,i}(q,p) = i g \frac{\lambda_i}{2} \Gamma_\nu(q,p)$. Here 
$p$ and $q$ denote the momenta of the two quark legs, whereas $k=p-q$ is the 
corresponding gluon momentum. The bare quark propagator is given by 
$[S^0(p)]^{-1} = i \gamma \cdot p + m(\Lambda^2)$, where $m(\Lambda^2)$ is 
the unrenormalised bare quark mass and $\Lambda$ represents an ultraviolet 
cutoff. The wave function renormalisation factor $Z_2$ is determined in the 
renormalisation process. The ghost renormalisation factor $\widetilde{Z}_3$ 
will be absorbed in our truncation of the quark-gluon vertex which we discuss 
below. The quark mass function $M(p^2)$ and the wave function $Z_f(p^2)$ 
can be extracted from Eq.~(\ref{quarkDSE}) by suitable projections in 
Dirac-space. 

\vspace*{3mm}
\subsection{The Bethe-Salpeter equation for the pion on a torus}\label{sec:bse}

\vspace*{-3mm}
The homogeneous Bethe-Salpeter equation (BSE) for flavour non-singlet 
mesons on a torus can be written as 
\begin{widetext}
\beq
\G_{\al\ba}^{\pi}(p_i;P)=\frac{1}{L^3 T}\sum_{j} K_{\al\ba;\de\ga}(p_i,q_j;P)\left[S(q^+_j)
\G^{\pi}(q_j;P)S(q^-_j)\right]_{\ga\de}
\label{eq:bse2}
\eeq
\end{widetext}
where $K$ is the Bethe-Salpeter kernel. The momenta $q^+=q+\xi P$ and 
$q^-=q-(1-\xi)P$ of the quark constituents are chosen such that the total 
meson momentum is given by $P=q^+-q^-$. Here the momentum partitioning 
parameter $\xi=[0,1]$ reflects the arbitrariness in the relative momenta 
of the quark-antiquark pair. Since all observables are independent of 
$\xi$ we can choose $\xi = \frac{1}{2}$ without loss of generality. 
The flavour content of the meson is expressed through flavour matrices 
which are suppressed in \Eq{eq:bse2}. The Greek indices ($\al\ldots$) 
refer to colour and Dirac structure. The BSE is a parametric eigenvalue 
equation with discrete solutions $P^2=-M_n^2$ where $M_n$ is the mass 
of the bound-state. The lowest mass solution corresponds to the physical 
ground state. Since $P^2$ is negative, the momenta $q^{\pm}$ are 
necessarily complex in Euclidean space and so the quark propagator 
functions must be evaluated with complex argument. This leads to 
technical issues that are dealt with in subsection \ref{sec:num}.

In general the pion Bethe-Salpeter amplitude can be decomposed into four
different tensor structures $F_{1..4}$ according to
\begin{eqnarray}
\Gamma^\pi(p_i;P)\!\!&=&\!\!\!\! \gamma_{5}\Big[F_1(p_i;P)
-i\Pslash \,\,F_2(p_i;P)\label{pion}\\              
&&\hspace{-9mm}-i\pslash_i \left(p_i\cdot P\right)F_3(p_i;P)
-\left[\Pslash,\pslash_i\right]F_4(p_i;P)\Big]\, .\nonumber
\end{eqnarray}
The specific values for the momenta $p_i$ depend on our choice of 
boundary conditions for the quark fields. In the time direction the relative 
momenta $p_i$ of the pion constituents have to be antiperiodic as can be seen 
from the exact expression \cite{Maris:1997hd}
\beq
F_1(p_i;P) = \frac{B(p_i)}{f_\pi}\;, \label{pichiral}
\eeq
in the chiral limit, where $B$ is the scalar quark dressing function and 
$f_\pi$ the pion decay constant. With our choice of antiperiodic boundary 
conditions for the quark fields in all four direction we obtain the same 
boundary conditions for the $p_i$. The total momentum $P$ of the pion is 
fixed by the condition $-P^2=M_\pi^2$ on the momentum shell of the pion.

The (normalised) Bethe-Salpeter amplitudes of the pion $\G^{\pi}$ can be used to
determine the pion decay constant according to
\beq
f_\pi = \frac{3 Z_2}{M^2} \frac{1}{L^3 T} \!\sum_{j}
\tr \!\left[\Gamma^\pi(q_j,-P) \,S(q^+_j) \,\gamma_5 \,\Pslash\,S(q^-_j)\right]
\label{fpi}
\eeq
on our torus. Here the trace is over Dirac matrices, and $q^\pm=q \pm P/2$.
As for the normalisation condition for the pion amplitude we use a prescription
originally proposed by Nakanishi in Ref.~\cite{Nakanishi:1965zza} and put to
use for momentum dependent kernels in \cite{Fischer:2009jm}. 

\subsection{Solving the DSE for complex momenta}\label{sec:num}

As we have seen, in the pion BSE the quark propagators are required for complex momenta  
\begin{equation} 
p^\pm = p \pm \frac{P}{2}\,,
\end{equation}
in the rest frame $P=(0,0,0,iM_\pi)$ of the pion and real $p$. 
To evaluate the quark DSE at
these complex arguments it is convenient to determine the quark dressing functions 
$A(p)$ and $B(p)$ as functions of $p$ instead of the usual $p^2$. Numerically we
solve the quark DSE for fixed gluon propagator and quark-gluon vertex by a plain
fixed point iteration method. This iteration has to be carried out on the grid of
real parts $p_j$ of $p^\pm$ with fixed imaginary parts parametrised by $M_\pi$.
Thus for any given value  of $M_\pi$ the quark DSE can be solved self-consistently 
without requiring knowledge of the propagator at a different $M_\pi$. Of course, 
due to symmetries it is not necessary to iterate the DSE for the full grid of 
momenta $p^\pm_j$ but for a subset parametrised by the absolute value of the $p_j$,
their four components $(p_j)_4$ and the pion momentum $M_{\pi}$.

When solving the quark DSE, Fig.\ref{DSEs}, in the complex plane one 
encounters the following ambiguity: since analytic continuation is not 
unique anymore on the discretised Matsubara frequencies one obtains 
results in the complex plane which are dependent on the momentum routing 
in the dressing loop of the DSE. This ambiguity is dependent on the volume. 
It is small for volumes larger than $V=(2 \,\mbox{fm})^4$, {\it i.e.} in the 
region where volume effects are small anyhow. Below $V=(1.5 \,\mbox{fm})^4$, 
however, there are sizeable effects. In this work we choose a momentum 
routing where the complex momentum goes through the quark part of the loop, 
whereas the gluon part depends on the real loop momentum only. Other choices
probe the gluon propagator in the complex plane where no numerical results
are available so far. 

\subsection{Truncation scheme for the quark-gluon interaction}\label{sec:trunc}
\begin{figure}[t]
\includegraphics[width=\columnwidth]{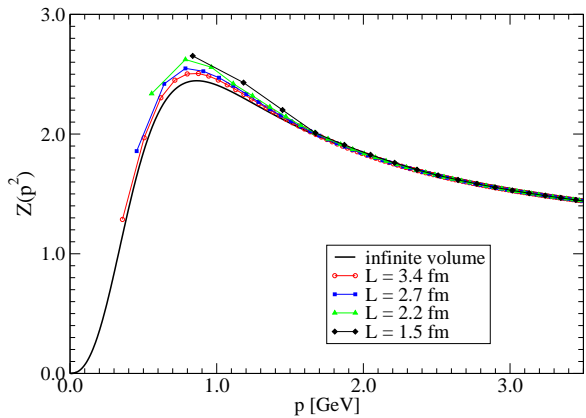}
\caption{Finite volume effects in the gluon dressing function $Z(p^2)$ 
as determined in Ref.~\cite{Fischer:2007pf}.}
\label{fig:glue}
\end{figure}

We now proceed by specifying the explicit interaction used in the 
quark-DSE Eq.~(\ref{quarkDSE}) and the BSE Eq.~(\ref{eq:bse2}). It consists 
of two distinct parts, the dressed gluon propagator and the dressed 
quark-gluon vertex. For the gluon propagator we use solutions from 
the ghost and gluon DSEs as obtained in 
Refs.~\cite{Fischer:2007pf,Fischer:2008uz}. There the system of DSEs 
has been solved using a specific truncation scheme for the ghost-gluon 
vertex and the three-gluon vertex as input. The resulting dressing 
functions for the ghost and gluon propagator are in good qualitative 
agreement with corresponding lattice results with a quantitative 
difference of the order of ten percent at the peak of the gluon 
dressing function for momenta around $p=1$ GeV. This is in marked 
contrast to corresponding dressing functions obtained from other 
approaches as {\it e.g.} the background gauge formalism 
\cite{Aguilar:2008xm}, where there is almost an order of magnitude 
difference between the characteristic scale of the DSE-solution and 
the lattice result. Thus, the results of \cite{Aguilar:2008xm} cannot 
be used in phenomenological calculations as they would necessitate 
an artificially strong quark-gluon vertex in order to provide for 
enough interaction strength to generate dynamical chiral symmetry 
breaking.

In general, two types of solutions for the ghost and gluon propagators
have been found in the deep infrared momentum region. There is a
`scaling' solution corresponding to an infrared vanishing gluon propagator
and an infrared diverging ghost dressing function and there is a continuous
family of `decoupling' solutions corresponding to an infrared finite
gluon propagator and an infrared finite ghost dressing function, see {\it e.g.}
\cite{Dudal:2007cw,Boucaud:2008ji,Fischer:2008uz} and references therein.
Recently, this family of decoupling solutions plus the limiting scaling 
one has been connected with ambiguities of fixing Landau gauge 
completely \cite{Maas:2009se}. It is furthermore a current issue of intense 
debate whether and how these two solutions relate to important fundamental 
questions like the mechanism of confinement of QCD and the possibility of 
a nonperturbative BRST-symmetry. Fortunately for the purposes of this 
work the behaviour in the deep infrared is irrelevant since scaling vs. 
decoupling distinguishes the ghost and gluon at momentum scales below 
$p^2=50$ MeV$^2$ corresponding to volumes far larger than the ones 
investigated here. 
\begin{figure}[t]
\centerline{\includegraphics*[width=\columnwidth]{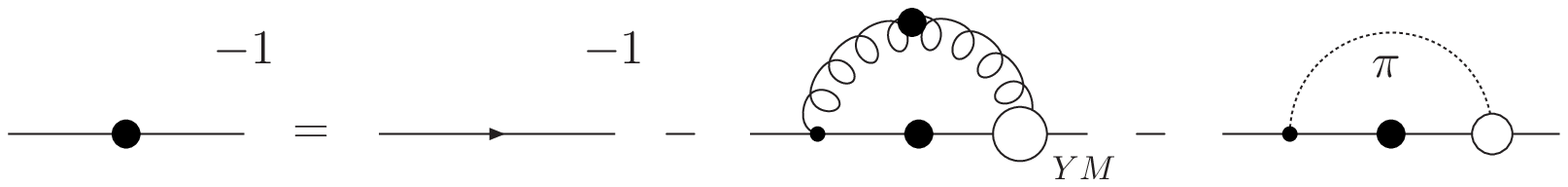}}
\caption{The approximated Schwinger-Dyson equation for the 
quark propagator with effective one-gluon exchange and 
one-pion exchange. 
\label{fig:quarkdse2}}
\vspace*{5mm}
\centerline{\includegraphics*[width=\columnwidth]{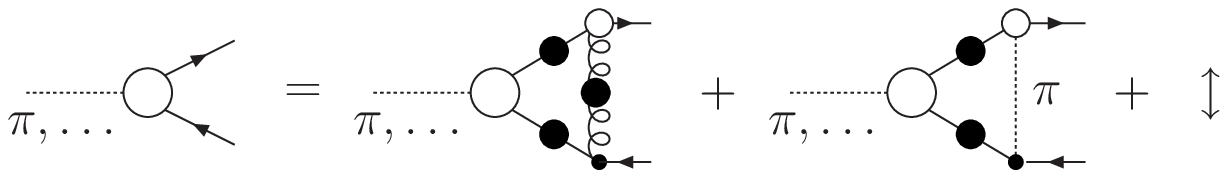}}
\caption{The Bethe-Salpeter equation corresponding to the 
quark self-energy of Fig.~\ref{fig:quarkdse2}. The up-down arrow indicates 
an averaging procedure of the pion exchange diagram with its counterpart
where the upper pion-quark vertex is bare and the lower one dressed.
\label{fig:BSE}}
\vspace*{5mm}
\end{figure}

For the gluon propagator at finite volume and discretisation we could in 
principle use results obtained on a torus \cite{Fischer:2007pf}, {\it cf.}
Fig.~\ref{fig:glue}. However, since 
corresponding results for the finite volume quark-gluon vertex are not yet 
available we believe it could be more systematic if we neglect volume effects 
in the gluon propagator as well. The numerical results in sections \ref{num:quark} 
and \ref{num:pi} are therefore obtained with the infinite volume/continuum 
gluon propagator evaluated at the bosonic Matsubara frequencies appearing 
in the quark DSE~\Eq{quarkDSE}. Nevertheless we also performed some calculations 
including finite volume effects in the gluon propagator which are discussed 
in section \ref{num:pi}.  

Next we give an explicit expression for the quark-gluon 
vertex. Here we follow the strategy of Ref.~\cite{Fischer:2008wy}
and split the vertex into a part containing only effects from pure Yang-Mills
theory and a part containing the effects of pion backreactions to the quark 
propagator. The resulting interaction is given diagrammatically in 
figures \ref{fig:quarkdse2} and \ref{fig:BSE}. It satisfies the axial-vector 
Ward-Takahashi identity which is mandatory to obtain the pion as a 
Goldstone boson in the chiral and infinite volume limit.

The Yang-Mills part of the quark gluon vertex in this approximation
is given by \cite{Fischer:2008wy}
\beqa
\Gamma_\nu^{YM}(k^2) &=& \gamma_\nu\, 
{Z_2}/{\widetilde{Z}_3} \, \, \Gamma^{\mathrm{YM}}(k^2) \,, \label{v1}\\
\Gamma^{\mathrm{YM}}(k^2) &=& \left(\frac{k^2}{k^2+d_2}\right)^{-1/2-\kappa} \label{model2}\\
&&\hspace{-1.5cm}\times\Bigg( \frac{ d_1}{d_2+k^2}+ \frac{ k^2 d_3}{d_2^2+\left(k^2- d_2\right)^2}
+\frac{k^2}{ { \lqcdsq}+k^2}
\nonumber \\[-1.2mm]
&&\hspace{-1.5cm}\times\left[\frac{4\pi}{\beta_0\alpha_\mu}
\left(\frac{1}{\log\left(\frac{k^2}{\lqcdsq}\right)}
-\frac{\lqcdsq}{k^2-\lqcdsq}\right)\right]^{-2\delta}
	    \Bigg) \nonumber
\eeqa
with the gluon momentum $k^2$, the one-loop value 
$\delta = \frac{-9 N_c}{44 N_c - 8 N_f}$ for the anomalous dimension of the 
re-summed ghost dressing function and $\alpha_\mu=0.2$. We also use 
$\lqcdsq=0.52$ GeV$^2$ similar to the scale obtained in Ref.~\cite{Alkofer:2003jj}. 
The infrared exponent $\kappa$ has been determined analytically in 
\cite{Zwanziger:2001kw,Lerche:2002ep} and is given by
$\kappa=\left(93-\sqrt{1201}\right)/98\simeq0.595$. While $Z_2$ is determined
selfconsistently in the quark-DSE we choose $\widetilde{Z}_3=1$ for the
ghost renormalisation factor. Since the 
combination of $\alpha_\mu$, the gluon dressing function and the two  
factors of $1/\widetilde{Z}_3$ from the bare and dressed quark-gluon vertex
together form a renormalisation group invariant this choice is possible without
affecting multiplicative renormalisability of the quark-DSE.
The only remaining free parameters of our interaction are $d_1$, $d_2$ and $d_3$.
These have been determined in \cite{Fischer:2008wy} to reproduce the
physical pion mass and decay constant as well as a reasonable mass of the $\eta'$
in the chiral limit \cite{Alkofer:2008et}. The resulting choice is 
$d_1=1.45$~GeV$^2$, $d_2=0.1$~GeV$^2$ and $d_3=3.95$~GeV$^2$.

The construction \Eq{v1} follows the frequently used rainbow-ladder
approximation of the full vertex and therefore involves only 
the $\gamma_\nu$-part of its tensor structure. Note, however,
that it has been shown in Ref.~\cite{Alkofer:2008tt} that such 
a model cannot capture all essentials of dynamical chiral symmetry 
breaking. Nevertheless it represents a useful starting point
for our investigation. 

It remains to specify the quark meson vertex for the backreaction of the
pion to the quarks. Here we employ only the leading part of the pion
amplitude \Eq{pion} given by the function $F_1(p_i;P)$. In Ref.~\cite{Fischer:2008wy}
the chiral limit relation \Eq{pichiral} has been used to represent the pion.
However, since $B \equiv 0$ in the chiral limit on a torus this is not an option here.
Instead we use the approximation
\beq
\Gamma^\pi(p_i;P) = \gamma_{5} F_1(\Re(p_i^2),p_i \cdot P=0,P^2)
\eeq
where $\Re(p_i^2)$ is the real value of $p_i^2$. Since $F_1$ is calculated from
the volume dependent pion BSE this approximation takes volume effects in the pion
wave function into account. As we will see, this model is elaborate enough to 
obtain a qualitative picture of volume effects due to the pion backreaction. 

\section{Numerical Results}\label{num}

Having discussed the details of our truncation for the quark-gluon interaction 
we now proceed to present our numerical results. All results shown are obtained 
using antiperiodic boundary conditions in all four space-time directions. 
We will also use $L=T$; other choices are possible in principle and will be dealt 
with elsewhere.

\subsection{Volume and quark mass dependence of the quark condensate \label{num:quark}}

\begin{figure*}[t]
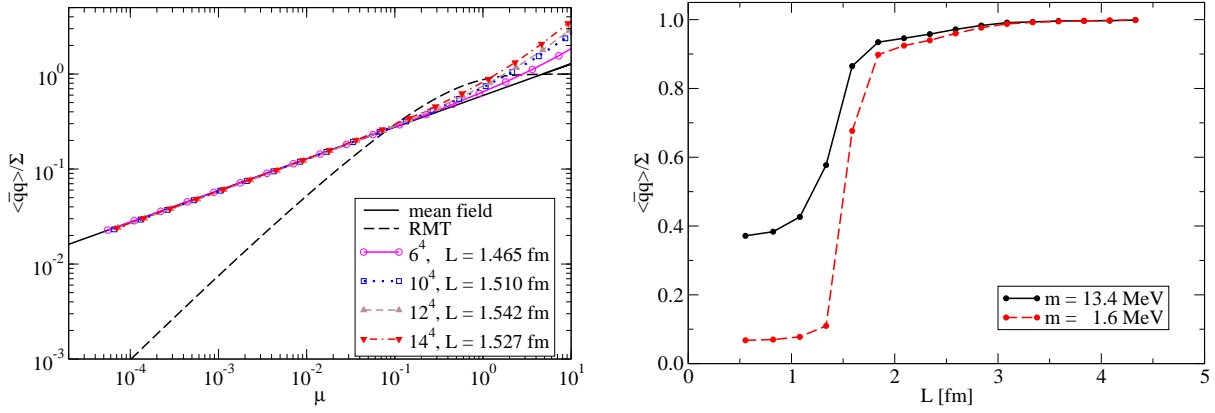

\includegraphics[width=\columnwidth]{mf.eps}\hfill
\includegraphics[width=\columnwidth]{cond3.eps}
\caption{Left diagram: The quark condensate $\langle \overline{q} q \rangle$ 
normalised by its infinite volume chiral value $\Sigma$ as a function of the 
dimensionless variable $\mu=mV\Sigma$. Shown are mean field scaling and 
the result from random matrix theory compared to our results from the 
quark-DSE for several lattice sizes
and box lengths.\\
Right diagram: The quark condensate evaluated as a function of box length $L$
for two different quark masses $m$ corresponding to pion masses of 
$M_\pi=100, 300$ MeV.}
\label{res:fig1}
\vspace*{0mm}
\end{figure*}

\begin{figure*}[t]
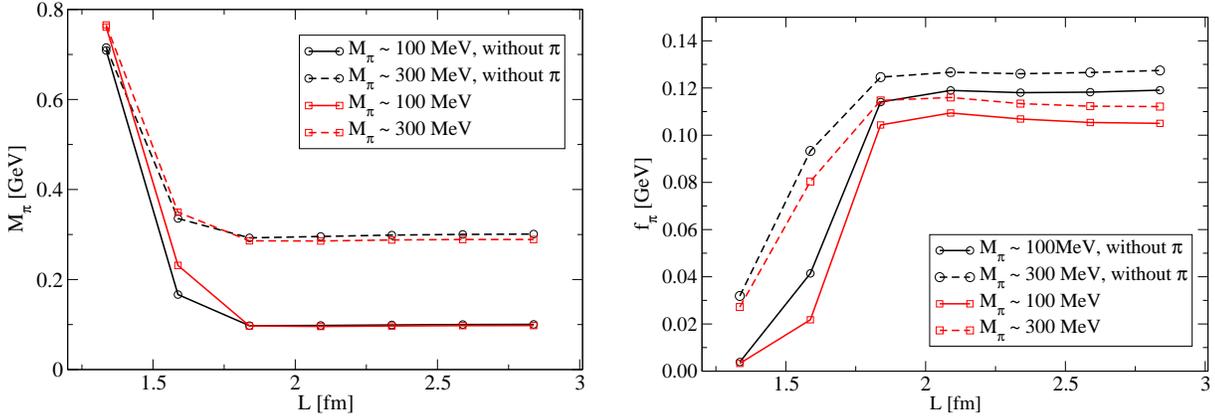

\includegraphics[width=\columnwidth]{mpi.eps}\hfill
\includegraphics[width=\columnwidth]{fpi.eps}
\caption{Left diagram: The pion mass as a function of box length $L$ for two
different infinite volume pion masses once without ('without $\pi$') and 
once including the pion backreaction to the quark propagator.\\
Right diagram: Same as left diagram but for the pion decay constant.}
\label{res:fig2}
\end{figure*}

Let us first discuss the volume and quark mass dependence of the 
quark condensate. With finite volume $V=L^3 T$ 
this quantity can be extracted from the dressed quark propagator via
\beq
\langle \overline{q} q \rangle =   \frac{4 Z_2 N_c}{L^3 T} \sum_{j}
\frac{B(p_j^2)}{p_j^2 A^2(p_j^2) + B^2(p_j^2)}\,, \label{cond}
\eeq
The result is a renormalisation point independent but cutoff dependent
condensate. At finite bare quark masses, necessary in a finite volume,
this quantity contains quadratic divergences in the continuum limit
of infinite ultraviolet cutoff. Therefore in principle one has to think
about the necessity of subtractions. However, it has been argued in 
Refs.~\cite{Leutwyler:1992yt,Damgaard:1999tk} that subtractions are 
not necessary in the finite volume scaling regime, {\it i.e.} in the 
region of volume and quark mass where the condensate is a function of 
the dimensionless variable $\mu=mV\Sigma$ only. Here $\Sigma$ represents 
the chiral condensate in the infinite volume limit. In this region the 
ultraviolet divergences of the form $m \Lambda^2$ and $m^3 \ln \Lambda$ 
with cutoff $\Lambda$ are suppressed by $1/V$ and $1/V^3$ respectively.
These corrections are non-universal.

The leading behaviour of $\langle \overline{q} q \rangle (\mu)$, however,
is universal in the scaling regime and can be calculated
from random matrix theory (RMT) \cite{Verbaarschot:2009jz}. For quarks 
in the fundamental representation of $SU(3)$ gauge theory the result is 
given by 
\beqa
\frac{\langle \overline{q} q \rangle}{\Sigma} 
&=& \mu \left[I_{N_f + \nu}(\mu) \, K_{N_f + \nu}(\mu) \right.\nonumber\\
&&\vspace*{1cm}\left. + I_{N_f + \nu + 1}(\mu) \, K_{N_f + \nu -1}(\mu)\right]
\label{scaling}
\eeqa
where the $I_n$ and $K_n$ are modified Bessel functions and $\nu$ is the 
topological charge.

\begin{figure*}[t]
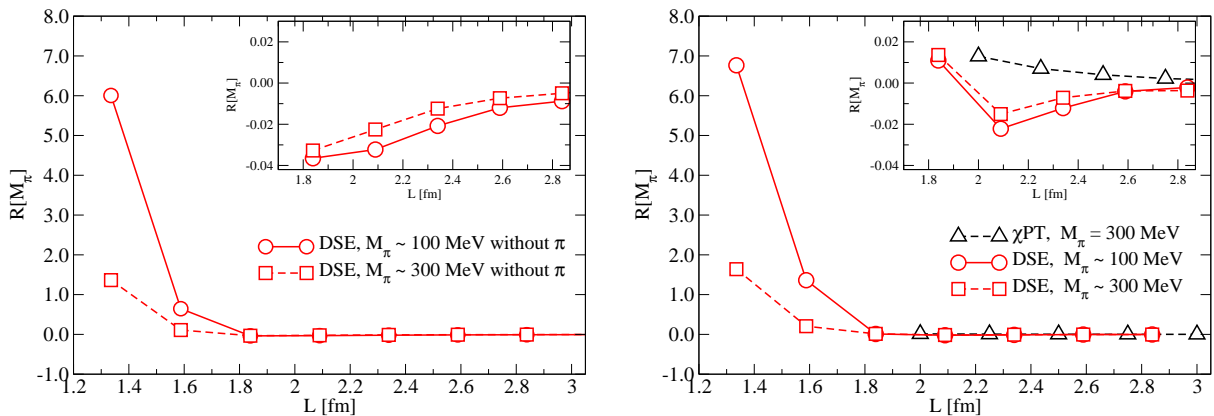

\includegraphics[width=\columnwidth]{Rpi.lin.quenched.eps}\hfill
\includegraphics[width=\columnwidth]{Rpi.lin.eps}
\caption{The $R[M_\pi]$ ratio as a function of box length. Shown are our
results for two different pion masses without (left diagram) and with
the inclusion of pion cloud effects (right diagram). The latter results
are compared to corresponding ones from chiral perturbation theory 
\cite{Colangelo:2005gd}. The inlays magnify the region where our results
have a negative sign.}
\label{res:fig3}
\end{figure*}

In the following we determine the quark condensate in this scaling regime
from solutions of our quark DSE for appropriate volumes and quark masses.
We work in quenched approximation, {\it i.e.} $N_f=0$. Our result is
displayed in the left diagram of Fig.~\ref{res:fig1} together with RMT 
scaling according to \Eq{scaling}. We clearly do observe scaling of the
condensate with the dimensionless variable $\mu$. However, the quantitative 
details of our results do not agree with the scaling behaviour predicted 
by RMT. So what is going wrong ? The reason for this discrepancy can be explained
from the limitations of our truncation scheme: of course one would expect to see 
RMT-scaling if and only if all Green's 
functions used in the quark-DSE would scale with $\mu$. Unfortunately this 
is not the case for our volume and quark mass independent input, namely 
the gluon propagator and the dressed quark-gluon vertex. These quantities 
constitute a $\mu$-independent background in which the quark condensate 
is evaluated. Nevertheless the volume and quark mass effects associated 
with the quark-DSE \Eq{quarkDSE} itself are present. As a result one 
should expect mean field scaling of the quark condensate as a function of $\mu$:
\beq
\langle \overline{q} q \rangle (\mu) \sim \mu^{1/\delta} \,, \label{MF}
\eeq
with $\delta=3$. This is indeed what we observe. As can be seen from 
Fig.~\ref{res:fig1}, in the scaling region $\mu \le 1$ our result satisfies 
\Eq{MF} with very good accuracy. We also observe critical slowing down 
in our numerics. We have checked that the scaling \Eq{MF} persists in 
our solutions regardless of the details of the truncation of the 
quark-gluon vertex and also for different boundary conditions of the 
quark field. Around $\mu \sim 1$ the aforementioned cut-off effects set 
in and take over completely for $\mu > 10$. It will certainly be 
interesting to check whether RMT-scaling appears in our approach when 
the restrictions on the quark-gluon vertex are lifted and volume and 
quark mass effects in the vertex are taken into account. This is, 
however, beyond the scope of the present work.

In the right hand diagram of Fig.~\ref{res:fig1} we also show the quark
condensate evaluated as a function of box length for two different quark
masses. Again we clearly observe that there is a critical region of 
volume below which the pattern of chiral symmetry breaking is changed
substantially and finally lost for very small volumes 
\cite{Fischer:2005nf,Fischer:2007ea}. This happens roughly around 
$L = 1.5$ fm which is also the characteristic length scale for the mean
field scaling we observe.

\subsection{Volume and quark mass dependence of the pion mass and decay constant \label{num:pi}}

In the previous section we discussed the volume dependence of a gauge invariant 
quantity, the quark condensate. Next we focus on two other gauge invariant
quantities, the pion mass and the pion decay constant. In Fig.~\ref{res:fig2} 
we show our results for the volume dependence of these quantities. In order to
also discuss effects due to different bare quark masses we include results 
corresponding to two different pion masses in the infinite volume limit, 
$M_\pi \approx 100, 300$ MeV. Furthermore in each diagram we show results with 
and without taking the pion backreaction to the quark propagator into account.

As concerns the pion mass, we see only very small effects for volumes larger 
than $V=(1.8 \mbox{fm})^4$. These volumes seem to be large enough to be 
close to the infinite volume limit. For the pion decay constant, however,
the situation is different: there are deviations from the infinite volume 
limit of the order of five percent up to volumes of $V=(2.3 \mbox{fm})^4$. 
Recalling \Eq{fpi} we find that the calculation
of $f_\pi$ directly includes the pion wave function. On the other hand this 
quantity is absent in the rainbow-ladder calculation of the pion mass and 
only present at a subleading level when we additionally backcouple the pions 
to the quarks. Our result for $f_\pi(L)$ then indicates that volume effects 
in the wave function of the pion are larger than the ones in its pole mass. 
This is as expected from the general considerations of Ref.~\cite{Leutwyler:1992yt}. 

In Fig.~\ref{res:fig3} we present results for the quantity
\beq
R[M_\pi] = \frac{M_\pi(L) - M_\pi(\infty)}{M_\pi(\infty)}
\eeq
which is sensitive to the deviation of the pion mass in a box of length 
$L$ compared to the infinite volume limit. In the left diagram we show results
without the inclusion of the pion backreaction on the quark propagator,
in the right diagram we show results taking these pion cloud effects
into account. We study two different pion masses $M_\pi=100, 300$ MeV.
In general the size of the volume effects are larger for the smaller pion
mass as expected. In both calculations we observe a positive shift of $R[M_\pi]$
if the box length is very small, followed by a turnover and a negative
shift for box lengths larger than $L \approx 1.6-1.9$ fm. The negative
shift is larger and sets in at smaller volumes if pion cloud effects are 
omitted (left diagram). Therefore it seems as if this negative shift may
predominantly be a property of the quenched theory. Indeed, the very same 
sign change in the $R[M_\pi]$-function has also been observed in quenched 
lattice QCD \cite{Guagnelli:2004ww}. Our results without pion backreaction 
are therefore in qualitative agreement with the lattice results. In order 
to compare also quantitatively we would have to introduce periodic boundary 
conditions in the spatial directions as well as different box lengths in the 
space and time directions. This study is relegated to future work. 

In the right diagram of Fig.~\ref{res:fig3} we compare our results 
(including pion cloud effects) with a corresponding one for the heavy 
pion extracted from a resummed form of L\"uscher's formula together 
with input from chiral perturbation theory at next-to-next-to-leading 
order \cite{Colangelo:2005gd}. For the light pion the $\chi PT$-calculation 
is outside the range of applicability of the $p$-counting-regime and therefore 
not shown in the plot. For the heavy pion mass our results show the onset 
of volume effects at similar length scales than $\chi PT$, however the
sign is different. $\chi PT$ predicts a positive shift in $R[M_\pi]$ for
all volumes. Since we have seen that our negative trend in $R[M_\pi]$
indeed gets smaller when we take the pion backreaction into account we can
not exclude that our result even changes into the positive domain when 
we treat the pion cloud effects more systematically. This will be explored
in future work. We wish to note, however, that negative values for $R[M_\pi]$
have also been observed in the quark meson model for unequal spatial and 
temporal box lengths, see \cite{Braun:2004yk,Braun:2005gy} for details.

\subsection{Additional volume effects due the gluon}\label{num:gluon}

\begin{figure}[t]
\includegraphics[width=\columnwidth]{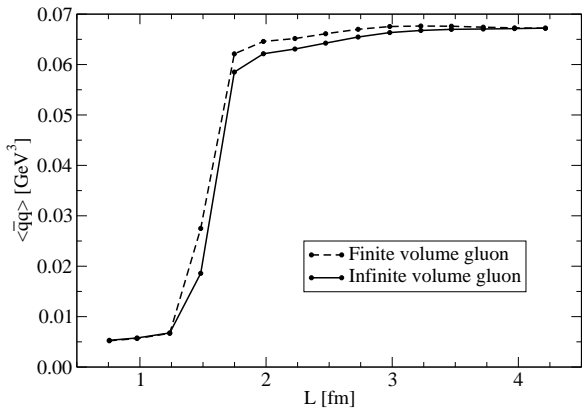}
\caption{The chiral condensate as a function of box length once calculated
with and once without finite volume effects in the gluon propagator.}
\label{res:fig4}
\end{figure}

Finally we discuss additional volume effects in the chiral condensate
due to volume effects in the gluon propagator. As discussed in section
\ref{sec:trunc} these effects should be included together with the
corresponding effects in the quark-gluon vertex. Nevertheless it may
be instructive to explore the size of the effects due to the gluon alone.
For the quark mass function these have already been discussed in 
Ref.~\cite{Fischer:2005nf,Fischer:2007ea}. Note that in \cite{Fischer:2005nf}
the volume effects in the gluon propagator have been overestimated and
consequently unrealistically large effects in the quark mass have been 
found. This finding has been corrected in \cite{Fischer:2007ea} using
the updated calculation of the finite volume gluon propagator from 
\cite{Fischer:2007pf}. Here we use a similar input, as detailed in 
section \ref{sec:trunc}. 

Our results are shown in Fig.~\ref{res:fig4}. We clearly observe a 
similar qualitative behaviour with and without the finite volume gluon. 
The quantitative differences between the two calculations are small 
({\it i.e.} below five percent) apart from a small region around $L = 1.5$ fm,
where the vanishing of chiral effects happens at slightly different scales.
For very small volumes both results become equal. This is easily understood
since in this region the quark condensate is dominated from the bare quark 
mass and dynamical effects are almost absent. We also performed 
calculations of the pion mass and decay constant using the finite volume
gluon propagator and found even smaller effects there. Therefore we agree
with the previous finding of~\cite{Fischer:2007ea}: the effect of finite volume 
corrections in the gluon propagator on physical observables is small.

\section{Summary and conclusions}\label{sum}

In this work we presented an exploratory study of volume effects in the
quark condensate, the pion mass and its decay constant determined from
a coupled set of Dyson-Schwinger (DSEs) and Bethe-Salpeter equations (BSEs). 
These equations are capable of connecting the large and small regions of momenta,
quark masses and volumes, and therefore provide an interesting tool to
determine these effects systematically. In our exploratory study we
concentrated on volume effects generated in the quark DSE and the pion 
BSE and neglected corresponding effects in the quark-gluon vertex. This 
deficiency introduces a volume independent scale. Consequently, at small 
volumes we do not observe the universal scaling of the condensate
as a function of $\mu=mV\Sigma$ predicted from random matrix theory. 
Instead we find critical scaling with mean field exponents at box lengths
of $L \approx 1.5$ fm.
In general we observe strong volume effects in all quantities for box lengths 
below $L=1.8$ fm. For the pion these effects are somewhat larger in the wave 
function than in its pole mass. As a consequence the effects in the pion 
decay constant $f_\pi$ are also larger than in the pion mass. In the region 
above $L=2$ fm we find volume effects with opposite sign than the ones 
determined by chiral perturbation theory. Whereas chiral perturbation theory 
predicts an increase of the pion mass we find a slight decrease for $L=(2-4)$ fm 
with the increase setting in only below $L=2$ fm. In the quenched theory 
this behaviour is in qualitative agreement with results from 
lattice QCD \cite{Guagnelli:2004ww}. For the unquenched theory, however, 
the persistence of this effect in contrast to chiral perturbation theory 
points towards deficiencies in our truncation scheme. We expect this 
problem to be cured by taking into account explicit volume effects in 
the quark-gluon vertex along the lines of recent work on the infinite 
volume case \cite{Alkofer:2008tt,Fischer:2009jm}. This will be the 
subject of future work.

\section*{Acknowledgments}
We are grateful to Jens Braun, Bertram Klein, Jens Mueller and 
Jan Pawlowski for helpful discussions. 
This work has been supported 
by the Helmholtz-University Young Investigator Grant No. VH-NG-332, 
and by the Helmholtz International Center for FAIR
within the LOEWE program of the State of Hesse.

\end{document}